\documentclass[twocolumn,showpacs,superscriptaddress,preprintnumbers,floats,aps,pra,10pt]{revtex4-1}
\usepackage{graphicx}
\usepackage{dcolumn}
\usepackage{amsmath}
\usepackage{amsfonts}
\usepackage{amssymb}
\usepackage{slashed}
\usepackage{subfigure}
\usepackage{bm}
\providecommand{\U}[1]{\protect\rule{.1in}{.1in}}
\pdfoutput=1

\begin{document}
\title{Linking little rip cosmologies with regular early universes}

\author{F. Contreras}
\altaffiliation{felipe.contreras@usach.cl}
\affiliation{Departamento de Matem\'aticas y Ciencia de la Computaci\'on, Universidad de Santiago de Chile, Las Sophoras 173, Santiago, Chile.}
\author{N. Cruz}
\altaffiliation{norman.cruz@usach.cl}
\affiliation{Departamento de F\'isica, Universidad de Santiago de Chile, \\
Avenida Ecuador 3493, Santiago, Chile.}
\author{E. Elizalde}
\altaffiliation{elizalde@ieec.uab.es}
\affiliation{Institut de Ci\`encies de l'Espai (ICE, CSIC), Carrer de Can Magrans s/n,
Campus UAB, 08193 Bellaterra (Barcelona), Spain.}
\affiliation{International Laboratory for Theoretical Cosmology, TUSUR University, 634050 Tomsk, Russia.}
\affiliation{Yukawa Institute for Theoretical Physics, Kyoto University, Kyoto 606-8502, Japan.}
\affiliation{Tomsk State Pedagogical University, TSPU, 634061 Tomsk, Russia.}
\author{E. Gonz\'alez}
\altaffiliation{esteban.gonzalezb@usach.cl}
\affiliation{Departamento de F\'isica, Universidad de Santiago de Chile, \\
Avenida Ecuador 3493, Santiago, Chile.}
\author{S. D. Odintsov}
\altaffiliation{odintsov@ieec.uab.es}
\affiliation{Institut de Ci\`encies de l'Espai (ICE, CSIC), Carrer de Can Magrans s/n,
Campus UAB, 08193 Bellaterra (Barcelona), Spain.}
\affiliation{International Laboratory for Theoretical Cosmology, TUSUR University, 634050 Tomsk, Russia.}
\affiliation{Instituci\'o Catalana de Recerca i Estudis Avan\c cats (ICREA), Barcelona, Spain.}

\date{\today}

\begin{abstract}
\textbf{{Abstract:}} Cosmological fluids with a Generalized Equation of State (GEoS) are here considered, whose corresponding EoS parameter $\omega$ describes a fluid with phantom behavior, namely  $\omega<-1$, but leading to universes free of singularities at any past or future, finite time. Thus avoiding, in particular, the Big Bang and the Big Rip singularities, the last one considered to be typical in phantom fluid models. More specifically, such GEoS fluid cosmologies lead to regular Little Rip universes. A remarkable new property of these solutions is proven here, namely that they avoid the initial singularity at early times; therefore, they are able to describe emergent universes. Solutions of this kind had been studied previously, but only either as late time or as early time solutions; never as solutions covering both epochs simultaneously. Appropriate conditions are proposed here that relate the Little Rip cosmologies with the initial regular universe, for the future and past regimes, respectively. This is done by taking as starting point the conditions under which a given scale factor corresponds to a Little Rip universe.
\vspace{0.5cm}
\end{abstract}
\maketitle

\section{Introduction}
Current cosmological observations have established that the expansion of the Universe is going through a late time accelerated phase. In the framework of general relativity, this acceleration could be produced by an exotic fluid, called dark energy, which necessarily has a negative pressure causing an overall repulsive behavior of gravity at large cosmological scales (see, e.g,~\cite{KhurshudyanR} -~\cite{OdintsovR} for some reviews). Another issue that emerges from the cosmological data is that the EoS of this fluid may be represented in the very simple form $\omega=P/\rho$ (although more complicated EoS are sometimes considered in the literature), where $\omega$ lies very close to $-1$, most probably being below $-1$. For example, the last Planck
results yield $\omega=-1.03\pm 0.03$ for a constant $\omega$ model and assuming a flat universe~\cite{Planck}. On the other hand, the nine years of WMAP survey in combination with CMB+BAO+$H_{0}$ measurements for the flat space case give $\omega=-1.073^{+0.090}_{-0.089}$, which in combination with SNe data yields $\omega=-1.084\pm 0.063$~\cite{WMAP}. Furthermore, A. Rest \textit{et al.}~\cite{Rest}, using the $1.5$ year
measurements of the Pan-STARRS1 project combined with BAO+CMB(Planck)+$H_{0}$ and assuming a flat universe, have found the value $\omega=-1.166^{+0.072}_{-0.069}$, which  is inconsistent with the value of $-1$ at the $2.3\sigma$ level. These results are indicating that a phantom behavior of the dark energy component cannot be ruled out from current cosmological data, rather on the contrary (see~\cite{Singh} -~\cite{Jassal}). As a consequence, if this possibility is taken seriously, an effective approach to describe phantom EoS is the inclusion of Generalized Equations of State (GEoS).

The study of GEoS for the main fluid component of the universe has already some history. They were inspired, to the best of our knowledge, in the
particular behavior of Friedmann models in inflationary scenarios. In order to extend the range of known inflationary behaviors, Barrow~\cite{Barrow} assumed that the matter stress has a pressure $p$ and density $\rho$ that are related by the following model EoS
\begin{equation}
P(\rho)=-\rho-B\rho^{\lambda},
\label{1}
\end{equation}
where $B$ and $\lambda$ are both constant, with $B\neq 0$. The standard EoS of a perfect fluid, $p=(B-1)\rho$, is recovered when $\lambda=1$. A variation of Eq.~(\ref{1}) was discussed by Mukherjee \textit{et al}~\cite{Mukherjee}, who considered the form
\begin{equation}
P(\rho)=A\rho-B\rho^{1/2},
\label{2}
\end{equation}
the case with $A=-1$, as well as other more general EoS fluids, having been studied in~\cite{Odintsov} and~\cite{Stefancic}. In these works, cosmological solutions of dark energy models with generalized fluids were analyzed, focusing in the future expansion of the universe. A late time behavior of a universe filled with a phantom dark energy component with an EoS given by Eq.~(\ref{1}) was investigated in~\cite{Paul} and~\cite{Paul1}, where the allowed values of the parameters $A$ and $B$ were constrained using $H(z)-z$ data, a model independent BAO peak parameter, and a cosmic parameter (WMAP7 data). It is interesting to note that cosmological solutions of GEoS, in particular the GEoS in (\ref{1}) and (\ref{2}), have been investigated in order to describe both the behavior of the very early and very late universe regimes. Nevertheless,  the very remarkable fact was nowhere pointed out, that solutions without a future singularity, such as little rip solutions, can also represent in the past perfectly regular solutions, corresponding to emergent or bouncing universes, and vice-versa. Furthermore, such effective fluid description is typical for modified gravity~\cite{Odintsov1}.

The main aim of this paper is to prove that some solutions, which until now have been discussed as late-time solutions or as early-time solutions, exclusively, but never as a solution in both regimes, can in fact give rise to perfectly valid solutions in both regions, not developing singularities in the past
neither at any finite future time. We obtain also the mathematical conditions required for a cosmological solution in order to avoid both past and future singularities. In particular, we will exhibit exact solutions, previously found in the literature for late or for early times only, and which indeed fulfill these
conditions. They will be proven to be regular at every time, except in the strict limit $t\rightarrow\infty$

The outline of the paper is as follows. In Section II we obtain the conditions to be satisfied in order to have solutions with a little rip behavior at late time and a regular behavior at earlier times. In Section III we discuss some solutions found in previous investigations, either as little rip solutions or as early-time regular solutions, under this new perspective and with a corresponding singularity analysis. We show how these solutions fulfill the conditions encountered in Section II and, consequently, how they can be extended to the future, or to the past (depending on the studied case). In other words, they can be promoted to cosmological solutions without singularities of any sort, neither in the past nor in the reachable future. In Sections IV and V, we constraint the free parameters of the models analyzed in Section III with the supernova Ia data from JLA, and we compare them to each other and with $\Lambda$CDM. Section VI is devoted to conclusions and to a final discussion of the new perspectives opened by these findings. Units where $8\pi G=c=1$ will be used throughout this work.

\section{Conditions for little-rip and regular early-time universes}
We will here obtain the conditions to be fulfilled by the scale factor $a(t)$ in order to avoid the initial singularity and late time singularities of the Big Rip type. With this in mind, we must first understand in detail the evolution of the behavior of the scale factor. To this end, we take advantage of the proof that appears, in much detail, in App. $A$, and which demonstrates that when the flat FLRW metric is considered with only one dominant fluid and an EoS with a parameter of state $\omega<-1/3$, then the only universes without singularities are those of the Bouncing and of the Emergent types.

Both the Bouncing and the Emergent universes need the scale factor $a(t)$ to behave as a convex or strictly convex function. This condition is obtained by imposing $\omega$ to satisfy $\omega<-1/3$. The difference between considering the conditions $\omega<-1$ and $\omega<-1/3$ lies in the possibility to have Big Rip type singularities in the future, because if $-1/3>\omega>-1$ then the Hubble parameter $H$ is a decreasing function and therefore the Big Rip phenomenon does not occur. Identically, in the presence of quintessence, singularities of the Big Rip type do not exist. However, as is well known, in presence of phantom matter it is indeed possible to have this type of singularities. 

In what follows, we will first recall the results previously found concerning the situations that lead to Little Rip models. Secondly, we will analyze the conditions in order to avoid past and future singularities, namely initial singularities and Big Rip ones.   We will then proceed to match both constraints and to obtain the general conditions that lead to regular universes. Finally, we  will  study the behavior of the EoS parameter $\omega$ when the density $\rho$ is equal to zero in the Bouncing models.

\subsection{Conditions for a Little Rip}
The conditions under which a dark energy density that increases with time, with EoS parameter $\omega<-1$, is able to avoid, in fact, a finite-time future singularity, were discussed in~\cite{Frampton}. Below there is a summary of the main conditions found.

Assuming that $a(t)$ can be written in the form
\begin{equation}
a(t)=e^{f(t)},
\label{4}
\end{equation}
the condition for this scale factor to be a non singular function for all $t$ is translated into a non singular function $f(t)$ (although this last could still tend to $-\infty$, as is rigorously stated in App. A). Then, the conditions $\omega<-1$ and $\rho>0$ lead to $d\rho/da>0$, which for $f(t)$ implies the following restriction
\begin{equation}
\ddot{f}>0.
\label{5}
\end{equation}
Thus, all Little Rip models with $\omega<-1$ are described by a scale factor given by an equation of the form (\ref{4}), with a non singular function $f$ satisfying  Eq.~(\ref{5}). In~\cite{Frampton} the conditions were also considered for the case of a Little Rip singularity, when both the EoS and the density as a function of the scale factor $\rho(a)$ are specified, but in our case we will only need to use the scale factor, $a(t)$.

It is important to mention that, also in~\citep{Frampton}, it was found that the above little rip solutions are consistent with the $\Lambda$CDM bounds, i. e., compliance with the supernovae data force the Little Rip model into a region of parameter space in which the model resembles $\Lambda$CDM.

\subsection{Conditions for Regular Universes}
In this subsection we will find the conditions for a universe that is regular at both early and late times. The conditions for a Little Rip only consider future times, while for those here we need, in addition, the early time conditions. Early singularities show up when the scale factor satisfies $a(t_{0})=0$, $a\xrightarrow{t\rightarrow t_{0}}-\infty$, or  $a\xrightarrow{t\rightarrow -\infty}0$, at some specific time $t=t_{0}$. Considering the early singularities and the convexity of the scale factor $a(t)$, it becomes possible to modify  Eq.~(\ref{4}) through a non-negative constant in such a way that the conditions for a Little Rip are included in the general conditions for a regular universe to exist at all values of $t$. Inspired in the Little Rip Eq.~(\ref{4}), let us consider the following scale factor
\begin{equation}
a(t)=\exp (g(t))+s,
\label{adem}
\end{equation}
where $g$ is chosen as a non singular function, in order to avoid both a singularity of the type $a\xrightarrow{t\rightarrow \pm t_{0}}\infty$ and one of the type $a(t_{0})=0$, and $s$ is a non-negative constant, in order to avoid  singularities of the kind $a\xrightarrow{t\rightarrow -\infty}0$. If $g\xrightarrow{t\rightarrow -\infty}-\infty$, then it is necessary to consider $s >0$ in order to get a universe with a minimum spatial size. In the other case, namely when the function $g(t)$ does not converge to $-\infty $, it is not necessary to consider a positive constant $s$, and it is allowed that $s$ may take the value zero. Therefore, Eq.~(\ref{adem}) represents a scale factor that avoids both types of singularities: the initial singularity and the late-time, Big Rip one.

It is compulsory to study now how the condition $\omega<-1$ leads to the corresponding conditions upon $g$ and $s$. In what follows, we will just consider a flat space in the FLRW metric. Let us start from the Friedmann equations
\begin{equation}
\left(\dfrac{\dot{a}}{a}\right)^{2}=\dfrac{\rho}{3},
\label{Fr1}
\end{equation}
\begin{equation}
\dfrac{\ddot{a}}{a}=-\dfrac{1}{6}\left(\rho+3P\right),
\label{Fr2}
\end{equation}
\begin{equation}
P=-2\dfrac{\ddot{a}}{a}-\left(\dfrac{\dot{a}}{a}\right)^{2},
\label{Fr3}
\end{equation}
the conservation equation
\begin{equation}
\dot{\rho }=-3\left(\dfrac{\dot{a}}{a}\right)\left(\rho+P\right),
\label{ct}
\end{equation}
and the one for the EoS parameter $\omega$,
\begin{equation}
\omega=\dfrac{P}{\rho}.
\label{parestado}
\end{equation}
In the case that there exists a point $t_{0}$ such that $P(t_{0})=\rho(t_{0})=0$, then the value of $\omega (t)$ at $t=t_{0}$ will be given by the following limit
\begin{equation}
\omega (t_{0})=\lim_{t\rightarrow t_{0}}{\frac{P(t)}{\rho(t)}}.
\end{equation}
In this way, using the restriction $\omega<-1$ and  Eq.~(\ref{ct}), we get that
\begin{equation}
\dfrac{d\rho}{da}=-\dfrac{3\rho}{a}(\omega+1)>0,\quad\text{for}\quad\rho\neq 0.
\label{drhoa}
\end{equation}
Since the only possible models are Bouncing or Emergent universes, then from Eq.~(\ref{Fr1}) one can see that $\rho$ reaches the value $0$ only at the bounce time $t_{b}$, in the Bouncing model. In the case of the Emergent universe, the scale factor is always growing. In this way, using Eqs.~(\ref{Fr1}), (\ref{Fr3}) and (\ref{parestado}), it becomes possible to rewrite  Eq.~(\ref{drhoa}), for the case
$\rho(t_{b})=0$ as
\begin{equation}
\begin{split}
&\dfrac{d\rho}{da}(t_{b})=\lim_{t\rightarrow t_{b}}-\dfrac{3\rho(t)}{a(t)}(\omega(t)+1) \\
& =\lim_{t\rightarrow t_{b}}-\dfrac{3}{a(t)}\left[3\left(\dfrac{\dot{a}(t)}{a(t)}\right)^{2}\right]\times \\
& \left(\left[-\dfrac{1}{3}-\dfrac{2}{3}\dfrac{a(t)\ddot{a}(t)}{\dot{a}(t)^{2}}\right]+1\right) \\
& =6\dfrac{\ddot{a}(t_{b})}{a(t_{b})^{2}}.
\end{split}
\end{equation}
Thus, the condition $\dfrac{d\rho}{da}>0$ must be fulfilled at all times except for the bouncing model if $\ddot{a}(t_{b})=0$. Therefore, in this case the inequality (\ref{drhoa}) becomes an equality at the bounce time $t=t_{b}$. Using  Eq.~(\ref{adem}) in Eq.~(\ref{Fr1}), we see that $\rho$ can be expressed as
\begin{equation}
\rho=3(\dot{a}/a)^{2}=3\left(\dfrac{\dot{g}e^{g}}{e^{g}+s}\right)^{2}.
\end{equation}
As a consequence, the condition $\dfrac{d\rho}{da}>0$ leads to
\begin{equation}
\dfrac{d\rho}{da}=6\dfrac{\left[\ddot{g}e^{2g}+\ddot{g}e^{g}s+\dot{g}^{2}e^{g}s\right]}{\left(e^{g}+s \right)^{3}}> 0,
\label{drhoa2}
\end{equation}
and, finally, we have from Eqs.~(\ref{drhoa}) and (\ref{drhoa2}) that $\omega<-1$ if and only if
\begin{equation}
e^{g}\ddot{g}+\ddot{g}s+\dot{g}^{2}s>0,
\label{constraintat}
\end{equation}
at all times, except for the time $t_{b}$ in the Bouncing model if $\ddot{a}(t_{b})=0$, in which case the inequality changes into an equality at the bounce time $t=t_{b}$. It is to be noted that the functions $a(t)$ in Eqs.~(\ref{4}) and (\ref{adem}) are very similar except for the constant $s$. Therefore, if $s=0$, the conditions for a Little Rip universe, and for one that is non-singular in the past, are the same as those obtained in Eq.~(\ref{constraintat}) for $s=0$. This is the case in most of the situations considered. Generally, it is  easier to represent $a(t)$ in terms of the function $g(t)$ only, after having taken the constant $s=0$.

In summary, the conditions for a  universe that is regular at all finite times with $s=0$ are 
\begin{equation}
\left\{ \begin{array}{cc}
\ddot{g}(t)>0, & t\neq t_{b},\\ \\
\ddot{g}(t)\ge 0, & t= t_{b}.
\end{array}
\right.
\label{crit1}
\end{equation}
In the other case, when the constant $s>0$, the conditions for getting a regular universe are
\begin{equation}
\left\{ \begin{array}{cc}
G(t) >0, & t\neq t_{b},\\ \\
G(t) \ge 0, & t= t_{b},  \\ \\
\end{array}
\right.
\label{crit2}
\end{equation}
where $G(t)=e^{g(t)}\ddot{g}(t)+\ddot{g}(t)s+\dot{g}(t)^{2}s$. The relation between the function $f$ in Eq.~(\ref{4}) and the function $g$ with the constant $s$ of  Eq.~(\ref{adem}), yields
\begin{equation}
g(t)=\ln \left( \exp(f(t))-s \right).
\end{equation}

Let us now analyze in detail what happens with the EoS parameter  $\omega $ when $\rho =0$.

\subsection{Criterion for Bouncing universes}
Let us suppose that we have a Bouncing universe dominated by a Phantom fluid with $\omega(t)<-1$. Moreover, assume that, at the time of the bounce, $t=t_{b}$, the relation $P(t_{b})\neq 0$ is satisfied. Then, using Eq.~(\ref{parestado}), we obtain that $\omega(t_{b})=-\infty $. Now, if $P(t_{b})=0$, then
Eq.~(\ref{Fr3}) leads to $\ddot{a}(t_{b})=0$. Furthermore, from Eqs.~(\ref{Fr1}), (\ref{Fr3}) and (\ref{parestado}), it comes out that $\omega(t)$ is given by
\begin{equation}
\omega(t)=\lim_{s\rightarrow t}-\dfrac{2}{3}\dfrac{a(s)\ddot{a}(s)}{\dot{a}\left(s\right)^{2}}-\dfrac{1}{3}.
\label{omeli}
\end{equation}

To know the value of $\omega(t_{b})$, a property that is considered in  App.~$B$ is used, which implies that if $f$ is an analytic function at $t_{0}$ such that
$f(t_{0})=\dot{f}(t_{0})=0$, then $\lim_{t\rightarrow t_{0}}\left|\dfrac{\dot{f}(t)}{f(t)}\right|=\infty $. Considering $\dot{a}$ as $f$ in the property of App.~$B$, and using $\ddot{a}\geq 0$ in Eq.~(\ref{omeli}), one reaches the conclusion that the behavior of $\omega $ in $t=t_{b}$ is given by
\begin{equation}
\omega(t_{b})=-\infty.
\end{equation}
This result not only says what happens with the EoS parameter $\omega $ at the time of the bounce, but also, and even more important, whether the scale factor represent a bouncing or an emergent model. The latter is because, when the scale factor $a$ represent an emergent universe, then the EoS parameter $\omega$ is a regular function. Therefore, if at the time of the bounce, $t_{b}$, $\omega $ satisfies $\omega(t_{b})=-\infty $, then the model is certainly a bouncing solution.

We conclude this section with an example that clearly illustrates this feature. Consider the following scale factor and its associated EoS parameter
\begin{equation}
\begin{array}{l}
a(t)=e^{t}\left[ 1+e^{-2t}\right], \\ \\
\omega(t)=-1-\dfrac{8e^{-2t}}{3(1-e^{-2t})^{2}}.
\end{array}
\end{equation}
This solution represents a Bouncing universe without singularities and with $\omega<-1$; nevertheless, there exists a point such that $\omega(t_{b})=-\infty$.

\section{Regular solutions of the GEoS}
In what follows, we will consider  cosmological Bouncing and Emergent solutions for a universe filled up with one exotic fluid, already studied before, and we will prove how these solutions satisfy the conditions indicated in Section II. Moreover, we show at the end of the section a figure that illustrates the initial regularity of the following five models.

\subsection{First solution}
To start, we restrict the GEoS of the fluid to a particular case of the general form given in Eq.~(\ref{1})
\begin{equation}
P(\rho)=-\rho-B\rho^{1/2}.
\label{7}
\end{equation}
The solution for the this GEoS with $B>0$ was first obtained in~\cite{Barrow} and was analyzed in~\cite{Cruz} as a regular solution at early time. The late behavior of this EoS yields a scale factor as a function of the cosmic time, given by
\begin{equation}
\begin{split}
a(t) & =a_{0}\exp\left(-\frac{2\rho_{0}^{1/2}}{3B}\right)\times \\
& \exp\left[\frac{2\rho_{0}^{1/2}}{3B}\exp\left(\frac{B\sqrt{3}}{2}(t-t_{0})\right)\right].
\end{split}
\label{13}
\end{equation}
It is necessary to mention that the double exponential behavior of this solution was previously found for a bulk viscous source in presence of an effective cosmological constant~\cite{Barrow1}. This is a consequence of the inclusion of bulk viscosity in the Eckart theory, which leads to a viscous pressure $\Pi$ of the type $-3\xi H$, where $\xi$ is usually assumed to have the form $\xi=\xi_{0}\rho^{\delta}$. In this solution, the asymptotic behavior of the scale factor is $a\xrightarrow{t\rightarrow\infty}\infty$, and $a\xrightarrow{t\rightarrow-\infty}a_{0}\exp\left(-\frac{2\rho_{0}^{1/2}}{3B}\right)$. The Hubble parameter is given by
\begin{equation}
H(t)=\frac{\rho_{0}^{1/2}}{\sqrt{3}}\exp\left[\frac{B\sqrt{3}}{2}(t-t_{0})\right],
\label{20}
\end{equation}
where the above expression indicates that $H$ is a positive function, with asymptotic behavior described by $H\xrightarrow{t\rightarrow\infty}\infty$ and,
$H\xrightarrow{t\rightarrow-\infty}0$.

\begin{figure}[ht]
\centering
\includegraphics[width=19pc]{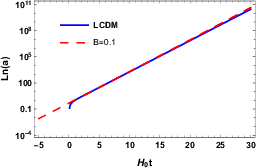}
\caption{\label{a-1}Plot of the scale factor as a function of $H_{0}t$ for $\Lambda$CDM and the first solution, using the last Planck result. For the first solution the initial condition $a_{0}=a_{\Lambda CDM}(3/H_{0})$ was used.}
\end{figure}

In Fig.\ref{a-1} we show a comparison for the scale factor of this model with $\Lambda$CDM, using the last Planck results ~\cite{Planck}. It is important to mention that when the parameter $B$ is close to $0$, the solution is reduced to $\Lambda$CDM for late times, as can be seen in Eq.(\ref{7}). In Fig.\ref{atot} we show the behavior of this solution at early times.

\subsubsection{Emergent and Little Rip solution}
As it was discussed in~\cite{Frampton} and briefly in  Sect. II.A here, the solution given in Eq.~(\ref{13}) can be written under the form $a=e^{f(t)}$, where in order to avoid the Big Rip, the function $f(t)$ must satisfy the condition $\ddot{f}(t)>0$. Owing to the fact that the minimum value of $a(t)$ is $a_{min}=a_{0}\exp\left(-\dfrac{2\rho_{0}^{1/2}}{3B}\right)>0$, it is possible to use Eq.~(\ref{adem}) taking $s =0$. In such case, the function $g(t)$ is equal to the function $f(t)$, and it can be represented by
\begin{equation}
\begin{array}{lc}
g(t)=&\dfrac{2\rho_{0}^{1/2}}{3B}\exp\left(\dfrac{B\sqrt{3}}{2}(t-t_{0})\right)\\
&+\ln (a_{0})-\dfrac{2\rho_{0}^{1/2}}{3B}.
\end{array}
\end{equation}
Its second derivative, $\ddot{g}(t)$, reads
\begin{equation}
\ddot{g}(t)=\dfrac{B\rho_{0}^{1/2}}{2}\exp\left(\dfrac{B\sqrt{3}}{2}(t-t_{0})\right),
\end{equation}
from where it is concluded that $\ddot{g}>0$. Therefore, using Eq.~(\ref{crit1}), it turns out that this scale factor leads to an Emergent regular universe dominated by a Phantom fluid. In order to check this fact, it is possible to study the behavior of the EoS parameter $\omega(t)$. Using  Eq.~(\ref{13}) in  Eq.~(\ref{omeli}), we get that $\omega(t)$ is given by
\begin{equation}
\omega (t)=-1-\dfrac{B\exp \left[ -\dfrac{1}{2}B\sqrt{3}(t-t_{0}) \right]}{\rho _{0}^{1/2}}.
\end{equation}
The above equation shows that $\omega<-1$ for all finite time, and that the scale factor represents an Emergent universe, because $\omega(t)$ is a regular function. Thus, it always has a phantom behavior and also a de Sitter like expansion at infinite future time.

\begin{figure}[ht]
\centering
\includegraphics[width=19pc]{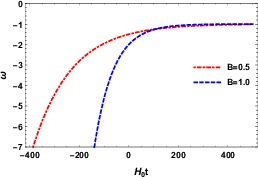}
\caption{\label{w-1}Plot of the parameter of state as a function of $H_{0}t$. Both curves are for $\rho_{0}=1$ and $t_{0}=0$. The values of the free parameters have been chosen, only to clearly expose the behavior of the parameter of state.}
\end{figure}

\subsection{Second solution}
In the introduction we already mentioned that Mukherjee \textit{et al.} took the GEoS given by Eq.(\ref{2}) and found for $A>-1$ and $B>0$ a scale factor of the form
\begin{equation}
a(t)=a_{i}\left(\beta+e^{\alpha t}\right)^{\gamma},
\label{amuk}
\end{equation}

where $a_{i}$ and $\beta$ are positive constants, $\alpha=B\sqrt{3}/2>0$, and $\gamma=2/3(A+1)$. 

The Hubble parameter is given by Eq.~(\ref{amuk}), as
\begin{equation}
H(t)=\frac{\alpha\gamma e^{\alpha t}}{\beta+e^{\alpha t}}.
\label{28}
\end{equation}

From these solutions it is not difficult to see that $a\xrightarrow{t\rightarrow\infty}\infty$, $a\xrightarrow{t\rightarrow-\infty}a_{i}\beta^{\gamma}$, $H\xrightarrow{t\rightarrow-\infty}0$, and $H\xrightarrow{t\rightarrow\infty}\alpha \gamma$.

\begin{figure}[ht]
\centering
\includegraphics[width=19pc]{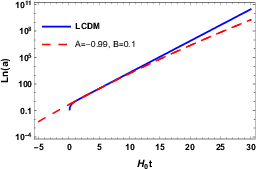}
\caption{\label{a-2}Plot of the scale factor as a function of $H_{0}t$ for $\Lambda$CDM and the second solution. For the second solution the initial condition $a_{0}=a_{LCDM}(3/H_{0})$ was used.}
\end{figure}

In Fig.\ref{a-2} we show a comparison for the scale factor of this model with $\Lambda$CDM. When the parameter $A$ is close to $-1$ and $B$ is close to $0$, the solution is reduced to $\Lambda$CDM for late times, as can be seen in Eq.(\ref{2}). In Fig.\ref{atot} we show the behavior of this solution at early times

\subsubsection{Emergent and Little Rip solution}
Owing to the fact that $H$ is always positive during the cosmic evolution, it is quite straightforward to see that, for $t\rightarrow\infty$, the solution behaves asymptotically as a de Sitter universe, with $H=\textup{const.}$, and that no future singularity exists. The minimum value of $a(t)$ is $a_{min}=a_{i}\beta ^{\gamma}$ and it is reached in the limit $t\rightarrow -\infty $. Therefore, we can use Eq.~(\ref{adem}), taking $s =0$, and obtain that the function $g(t)$ (which is now equal to the function $f(t)$), is given by
\begin{equation}
g(t)=\ln(a_{i})+\gamma\ln(\beta +e^{\alpha t}),
\end{equation}
while $\ddot{g}(t)$ reads
\begin{equation}
\ddot{g}(t)=\dfrac{\gamma \beta \alpha^{2}e^{\alpha t}}{\left(\beta +e^{\alpha t}\right)^2}.
\end{equation}
The above equation leads to the condition $\ddot{g}>0$. Therefore, using Eq.~(\ref{crit1}) one gets that this scale factor leads to an Emergent regular universe dominated by a phantom fluid. In order to check this, it is possible to study the behavior of the EoS parameter $\omega (t) $. Using Eq.~(\ref{amuk}) in Eq.~(\ref{omeli}), it turns out that
\begin{equation}
\omega (t)= -1-\dfrac{2\beta e^{-\alpha t}}{3\gamma}.
\end{equation}
The above equation shows that, in fact, $\omega <-1$ for all time and that the scale factor represents an Emergent universe, because $\omega (t)$ is a regular function.

\begin{figure}[ht]
\centering
\includegraphics[width=19pc]{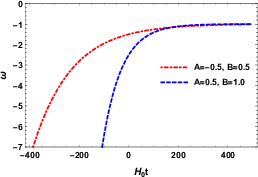}
\caption{\label{w-2}Plot of the parameter of state as a function of $H_{0}t$. Both curves are for $\beta=1$. The values of the free parameters have been chosen, only to clearly expose the behavior of the parameter of state.}
\end{figure}

\subsection{Third solution}
A cosmological inhomogeneous GEoS of the form
\begin{equation}
P(\rho)=-\rho-f(\rho)-\xi(H),
\label{31}
\end{equation}
was introduced in~\cite{Brevik} (for general review of viscous cosmology see~\cite{Brevik1}). The function $f(\rho)$ is an arbitrary one and $\xi(H)$ is a general function of $H$ related to the effective pressure for a fluid with viscosity. In the simple case $\xi(H)=\xi_{0}=$const., and taking $f(\rho)=B\rho^{1/2}$, the GEoS becomes
\begin{equation}
P(\rho)=-\rho-B\rho^{1/2}-\xi.
\label{eevis2}
\end{equation}

For a flat universe, an exact solution was found in~\cite{Brevik} in terms of the energy density as a function of time. Using this expression, it is now quite straightforward to see that an infinite time is needed to reach an infinite energy density, which correspond to a Little Rip under viscous conditions. As we are interested in exploring the behavior of this solution in more detail, we will integrate Friedmann's equations in order to find the explicit form of the scale factor as a function of the cosmic time. Using the GEoS given in Eq.~(\ref{eevis2}) in the continuity equation, Eq.~(\ref{ct}), and integrating using the initial conditions, $a(t=t_{0})=a_{0}$ and $\rho(t=t_{0})=\rho_{0}$, we obtain
\begin{equation}
\begin{split}
a(\rho)= & a_{0}\left(\frac{B\rho_{0}^{1/2}+\xi}{B\rho^{1/2}+\xi}
\right)^{\frac{2\xi}{3B^{2}}}\times \\
& \exp\left[\frac{2}{3B}\left(\rho^{1/2}-\rho_{0}^{1/2}\right)\right].
\end{split}
\label{33}
\end{equation}

From Eqs.~(\ref{eevis2}), (\ref{Fr1}) and (\ref{ct}) with the same initial conditions as for the scale factor, we obtain the energy density as a function of the cosmic time, as
\begin{equation}
\rho(t)^{1/2}=\frac{\exp\left[\frac{B\sqrt{3}}{2}(t-t_{0})\right]\eta-\xi}{B},
\label{34}
\end{equation}
where $\eta=B\rho_{0}^{1/2}+\xi$, which allows us to introduce in Eq.~(\ref{33}) the scale factor as a function of time
\begin{equation}
\begin{aligned}
a(t)= & \exp\left\{\frac{2}{3B^{2}}\left[\exp\left(\frac{B\sqrt{3}}{2}(t-t_{0})\right)\eta-\eta\right]\right\} \\
& \times a_{0}\exp\left[-\frac{\xi\sqrt{3}}{3B}(t-t_{0})\right],
\label{35}
\end{aligned}
\end{equation}
and a Hubble parameter
\begin{equation}
H(t)=\frac{\eta}{B\sqrt{3}}\exp\left[\frac{B\sqrt{3}}{2}(t-t_{0})\right]-\frac{\xi}{B\sqrt{3}}.
\label{36}
\end{equation}
From Eqs.~(\ref{35}) and (\ref{36}) it is possible to compute the asymptotic behavior of $a$ and $H$ as $a\xrightarrow{t\rightarrow\infty}\infty$, $a\xrightarrow{t\rightarrow-\infty}\infty$, $H\xrightarrow{t\rightarrow\infty}\infty$, and $H\xrightarrow{t\rightarrow-\infty}-\frac{\xi}{B\sqrt{3}}$.

\begin{figure}[ht]
\centering
\includegraphics[width=19pc]{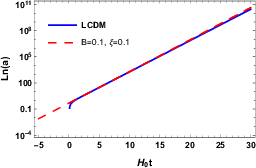}
\caption{\label{a-3}Plot of the scale factor as a function of $H_{0}t$ for $\Lambda$CDM and the third solution. For the third solution the initial condition $a_{0}=a_{LCDM}(3/H_{0})$ was used.}
\end{figure}

In Fig.\ref{a-3} we show a comparison for the scale factor of this model with $\Lambda$CDM. When the parameters $B$ and $\xi$ are close to $0$, the solution is reduced to $\Lambda$CDM for late, as can be seen in Eq.(\ref{eevis2}). In Fig.\ref{atot} we show the behavior of this solution at early times.

\subsubsection{Bouncing and Little Rip solution}
As it was pointed out in~\cite{Brevik}, this solution leads to a Little Rip for late times, which is straightforward to see from Eqs.~(\ref{35}) and (\ref{36}). In order to check our previous theorem, it is needed that the function $f(t)$ satisfies $\ddot{f}(t)>$, where $f(t)$ is given by $a=e^{f(t)}$. The minimum value of $a(t)$ is obtained from Eq.~(\ref{35}), as $a_{min}=a_{0}\left(\frac{\xi}{\eta}\right)^{-\frac{2\xi}{3B^{2}}}\exp\left(-\frac{2\rho_{0}^{1/2}}{3B}\right)$ when $t\rightarrow t_{b}=t_{0}+\frac{2}{B\sqrt{3}}\ln\left(\frac{\xi}{\eta}\right) $. Thus, we can use Eq.(\ref{adem}) considering $s =0$. In this case, we obtain that the function $g(t)$ (here again $g=f$) is given by
\begin{equation}
\begin{array}{ll}
g(t)=&\ln (a_{0})-\dfrac{\xi \sqrt{3}}{3B}(t-t_{0}) -\dfrac{2\eta}{3B^{2}} \\ \\
& + \dfrac{2\eta}{3B^{2}} \exp( \dfrac{B\sqrt{3}}{2}(t-t_{0})),
\end{array}
\end{equation}
and $\ddot{g}(t)$ reads
\begin{equation}
\ddot{g}(t)=\dfrac{\eta}{2}\exp \left[ \dfrac{B\sqrt{3}}{2}(t-t_{0})\right],
\end{equation}
where from it is clear that $\ddot{g}>0$. Therefore, using Eq.~(\ref{crit1}) the result is that this scale factor leads to a bouncing regular universe dominated by a phantom fluid. As this scale factor represents a bouncing universe with an EoS parameter $\omega<-1$, it is possible to use the above criterion for a bouncing universe, what exhibits that this effective EoS parameter $\omega$ reaches the value of $-\infty $ at the time of bounce $t=t_{b}$. Indeed, using Eq.~(\ref{35}) in Eq.~(\ref{omeli}), the effective $\omega(t)$ is obtained as
\begin{equation}
\omega(t)=-1-\dfrac{B^{2}\exp\left[\dfrac{B\sqrt{3}}{2}(t-t_{0})
    \right]}{\left(\eta\exp\left[\dfrac{B\sqrt{3}}{2}(t-t_{0})\right]-\xi\right)^{2}},
\end{equation}
where it is shown that $\omega <-1$ for all time and that the scale factor represents a bouncing universe, owing to the fact that $\omega (t_{b})=-\infty$.

\begin{figure}[ht]
\centering
\includegraphics[width=19pc]{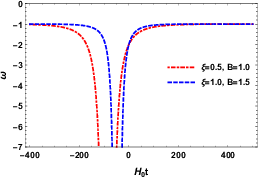}
\caption{Plot of the parameter of state as a function of $H_{0}t$. Both curves are for $\rho_{0}=1$ and $t_{0}=0$. The values of the free parameters have been chosen, only to clearly expose the behavior of the parameter of state.}
\label{w-3}
\end{figure}

\subsection{Fourth solution}
A scale factor with exponential behavior was studied in~\cite{Myrzakulov}, with the exact form
\begin{equation}
a(t)=a_{0}e^{\alpha (t-t_{0})^{2n}},
\label{a1Myr}
\end{equation}

and Hubble parameter
\begin{equation}
H(t)=2n\alpha(t-t_{0})^{2n-1}.
\label{H4}
\end{equation}
Here $a_{0}$ and $\alpha $ are positive constants, and $n$  a non-zero natural number, which affects the features of the bouncing. When $n<1/2$ or $n$ is a positive non-natural number, the bounce is changed into a finite-time singularity, occurring at $t = t_{0}$, or into an Emergent universe, with $a=0$ for $t\rightarrow -\infty $.

\begin{figure}[ht]
\centering
\includegraphics[width=19pc]{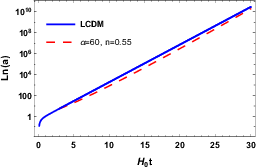}
\caption{\label{a-4}Plot of the scale factor as a function of $H_{0}t$ for $\Lambda$CDM and the fourth solution. For the fourth solution the initial condition $a_{0}=a_{LCDM}(3/H_{0})$ was used.}
\end{figure}

In Fig.\ref{a-4} we show a comparison for the scale factor of this model with $\Lambda$CDM. When the parameter $\alpha$ is close to $H_{0}$ and $n$ is close to $0.5$, the solution is reduced to $\Lambda$CDM for late times, as can be seen in Eq.(\ref{H4}). In Fig.\ref{atot} we show the behavior of this solution at early times.

\subsubsection{Bouncing and Little Rip solution}
Eq.~(\ref{a1Myr}) exhibits a scale factor $a(t)$ that corresponds to a Little Rip universe for late times, when $n$ is a positive natural number. To check this behavior, just as it was done before, it is necessary that the function $f(t)$ fulfills the condition $\ddot{f}(t)>0$, where $f(t)$ is given by the relation $a=e^{f(t)}$. The time of bounce is reached at $t=t_{b}=t_{0}$, with a scale factor taking the value of $a(t_{0})=a_{b}=a_{0}$. With this scale factor, it is possible to study when the EoS parameter  satisfies $\omega <-1$, and its behavior in the vicinity of the bounce point $t_{0}$. Since the constant $a_{0}>0$, it is possible to represent the scale factor (\ref{a1Myr}) by Eq.~(\ref{adem}) with $s=0$. In this case, one gets that the function $g$ (here $g=f$) is given by
\begin{equation}
g(t)=\ln{(a_{0})}+\alpha(t-t_{0})^{2n}, \quad n\in \mathbb{N}.
\end{equation}

The regularity of $g$ and its derivatives comes from the fact that $n\in\mathbb{N}$. Now, the function $\ddot{g}(t)$ reads
\begin{equation}
\ddot{g}(t)=\alpha 2n(2n-1)(t-t_{0})^{2(n-1)}, \quad n\in \mathbb{N}.
\end{equation}
The above equation reveals that $\ddot{g}>0\quad\forall n\in\mathbb{N}$ and $t\in\mathbb{R}$. Therefore, the EoS parameter $\omega$ satisfies $\omega<-1$. Indeed, using Eq.~(\ref{a1Myr}) in Eq.~(\ref{omeli}), we conclude that
\begin{equation}
\omega (t)=-1-\dfrac{(2n-1)}{3n\alpha (t-t_{0})^{2n}}.
\end{equation}
This equation corresponds to $\omega<-1$ for all times, and the scale factor represents a bouncing universe, due to the fact that $\omega =-\infty $ at
$t=t_{b}=t_{0}$.

\begin{figure}[ht]
\centering
\includegraphics[width=19pc]{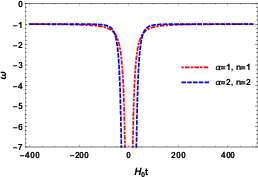}
\caption{\label{w-4}Plot of the parameter of state as a function of $H_{0}t$. Both curves are for $t_{0}=0$. The values of the free parameters have been chosen, only to clearly expose the behavior of the parameter of state.}
\end{figure}

\subsection{Fifth solution}
A power-law behavior for the scale factor was also studied in~\cite{Myrzakulov}, yielding in this case the exact form
\begin{equation}
a(t)=a_{0}+\alpha(t-t_{0})^{2n},
\label{a2Myr}
\end{equation}

\begin{figure}[ht]
\centering
\includegraphics[width=19pc]{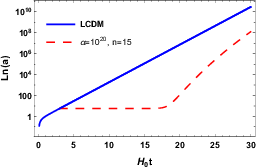}
\caption{\label{a-5}Plot of the scale factor as a function of $H_{0}t$ for $\Lambda$CDM and the fifth solution. For the fifth solution the initial condition $a_{0}=a_{LCDM}(3/H_{0})$ was used.}
\end{figure}

\noindent and a Hubble parameter
\begin{equation}
H(t)=\dfrac{2n\alpha(t-t_{0})^{2n-1}}{a_{0}+\alpha(t-t_{0})^{2n}},
\end{equation}
where $a_{0}$ and $\alpha $ are positive  constants, and $n$ is a positive natural number.

\begin{figure}[ht]
\centering
\includegraphics[width=19pc]{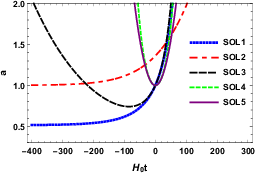}
\caption{\label{atot}Plot of the scale factors of these five models.  The value of the free parameters has been chosen only to clearly expose the behavior of these scale factors.}
\end{figure}

In Fig.\ref{a-5} we show a comparison for the scale factor of this model with $\Lambda$CDM. In this case the scale factor of the model cannot be matched with $\Lambda$CDM at late late times because this solution is a polynomial function, while in $\Lambda$CDM model the scale factor at late times is an exponential function. In Fig.\ref{atot} we show the behavior of this solution at early times.

\subsubsection{Bouncing and Little Rip solution}
Eq.~(\ref{a2Myr}) shows a scale factor $a(t)$ that corresponds to a bouncing regular universe. Now, we are going to check the Little Rip behavior from the condition $\ddot{f}(t)>0$. The time of bounce is reached at $t=t_{b}=t_{0}$, with a scale factor taking the value of $a(t_{0})=a_{b}=a_{0}$. With this scale factor, it is possible to study when $\omega <-1$ and the behavior of $\omega $ in the vicinity of the bounce point $t_{0}$. As  $a_{0}>0$, it is possible to represent the scale factor (\ref{a2Myr}) by Eq.~(\ref{adem}) with $s=0$. Thus, the function $g$ (again $g=f$) is obtained as
\begin{equation}
g(t)= \ln\left(a_{0}+\alpha(t-t_{0})^{2n}\right),
\end{equation}
and the function $\ddot{g}(t)$ can also be calculated for the above equation, and has the following form

\begin{figure}[ht]
\centering
\includegraphics[width=19pc]{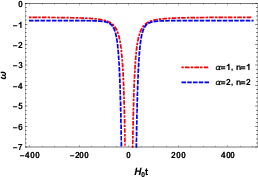}
\caption{\label{w-5}Plot of the parameter of state as a function of $H_{0}t$. Both curves are for $a_{0}=1$ and $t_{0}=0$. The values of the free parameters have been chosen, only to clearly expose the behavior of the parameter of state.}
\end{figure}

\begin{equation}
\ddot{g}(t)=\dfrac{2\alpha(t-t_{0})^{2n-1}[a_{0}(2n-1)-
    \alpha(t-t_{0})^{2n}]}{\left[a_{0}+\alpha(t-t_{0})^{2n}\right]^{2}}.
\end{equation}

In this case, the sign of the function $\ddot{g}$ depends of the value of $a_{0}$. This is due to the fact that the domain where $\omega<-1$ depends on the value of $a_{0}$. Using the above equation, one gets  that $\ddot{f}(t)>0$ for $t\in ( t_{0}-t_{s},t_{0}+t_{s})$, with $t_{s}=\left[ (2n-1)a_{0}/\alpha\right]^{1/2n}$. Therefore, $\omega<-1$ if $t\in ( t_{0}-t_{s},t_{0}+t_{s})$. It is possible to check this result by using Eq.~(\ref{omeli}) in order to obtain $\omega(t)$, with the result
\begin{equation}
\omega(t)=-1-\dfrac{a_{0}(2n-1)-\alpha(t-t_{0})^{2n}}{3\alpha n(t-t_{0})^{2n}}.
\end{equation}
We see that the EoS parameter $\omega $ at the time of bounce is $\omega(t_{0})=-\infty$. Moreover, $\omega<-1$ for $t\in (t_{0}-t_{s},t_{0}+t_{s})$, as is clear from the previous expressions.

\section{Constraining the models to supernova IA data}
Now we shall constraint the respective free parameters of the first three Little Rip models studied in the above section with the observational supernova Ia (SNe Ia) data. We do not constrain the fourth and fifth solutions, because these solutions are set by hand and do not have any relevant physics behing, like an equation of state that originates them. These solutions are being considered in this paper only as explicit proofs of the linking between the two domains, as discovered here. To impose the constraint, we use here the \textit{Joint Light curve Analysis} (JLA) sample (see~\cite{Betoule}), which contains 740 SNe up to redshift $z\backsimeq  1.3$, coming from nine different surveys.

\begin{figure*}
\centering
\includegraphics[scale=0.72]{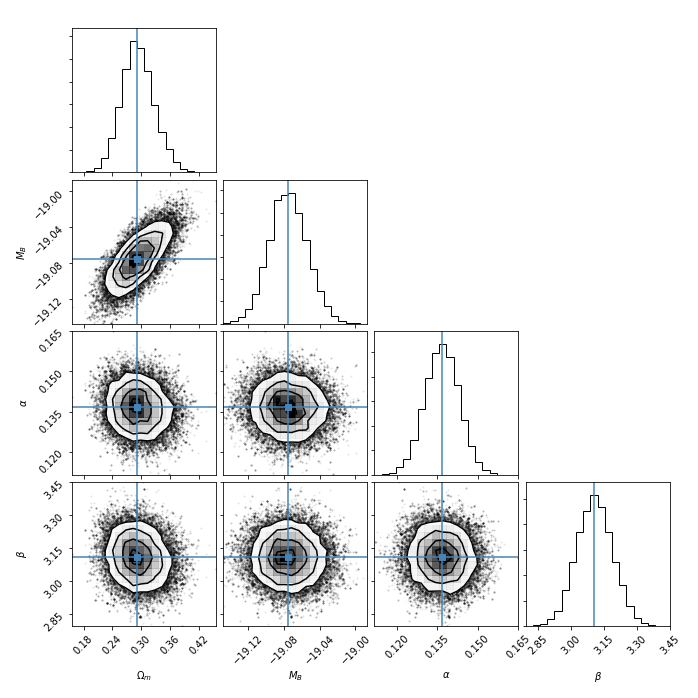} 
\caption{\label{trianglelcdm}Joint and marginalized constraint of $\Omega_{m}$, for the $\Lambda$CDM model, and marginalized constraint of the light-curve parameters $M_{B}$, $\alpha$ and $\beta$ of the JLA sample. The admissible regions correspond to $1\sigma(68.3\%)$, $2\sigma(95.5\%)$ and $3\sigma(99.7\%)$ confidence level (CL), respectively. The best fit values for each parameter are shown in Table \ref{Table I}.}
\end{figure*}

\begin{figure*}
\centering
\includegraphics[scale=0.59]{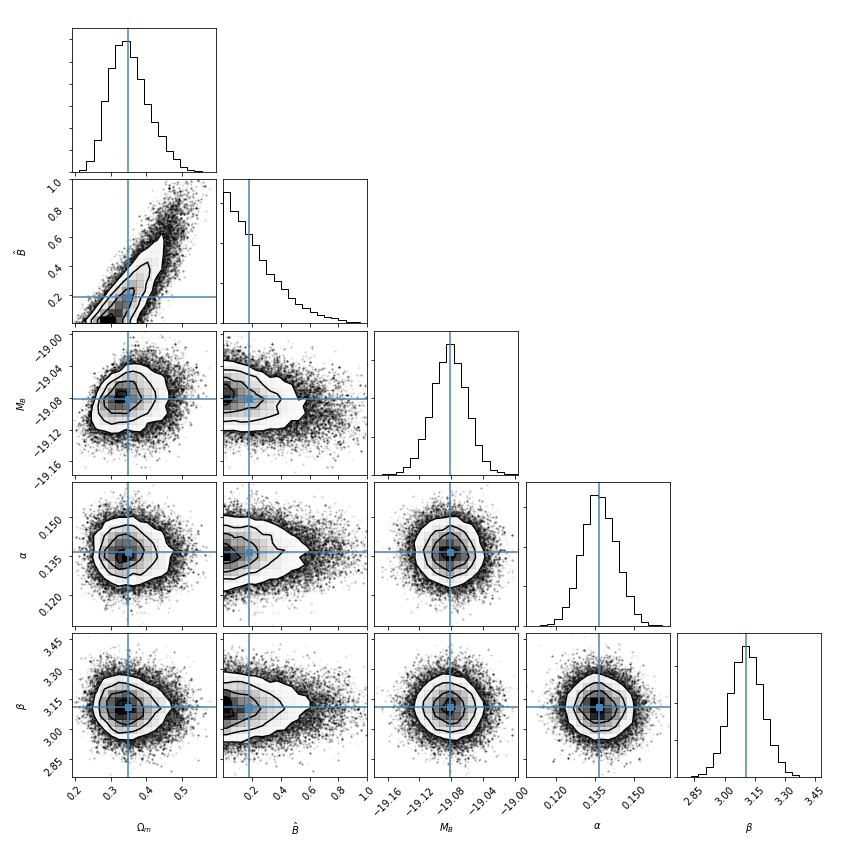} 
\caption{\label{triangle1}Joint and marginalized constraint of $\Omega_{m}$ and $\hat{B}$, for the model presented in the first solution, and marginalized constraint of the light-curve parameters $M_{B}$, $\alpha$ and $\beta$ of the JLA sample. The admissible regions correspond to $1\sigma(68.3\%)$, $2\sigma(95.5\%)$ and $3\sigma(99.7\%)$ confidence level (CL), respectively. The best fit values for each parameter are shown in Table \ref{Table I}.}
\end{figure*}

\begin{figure*}
\centering
\includegraphics[scale=0.50]{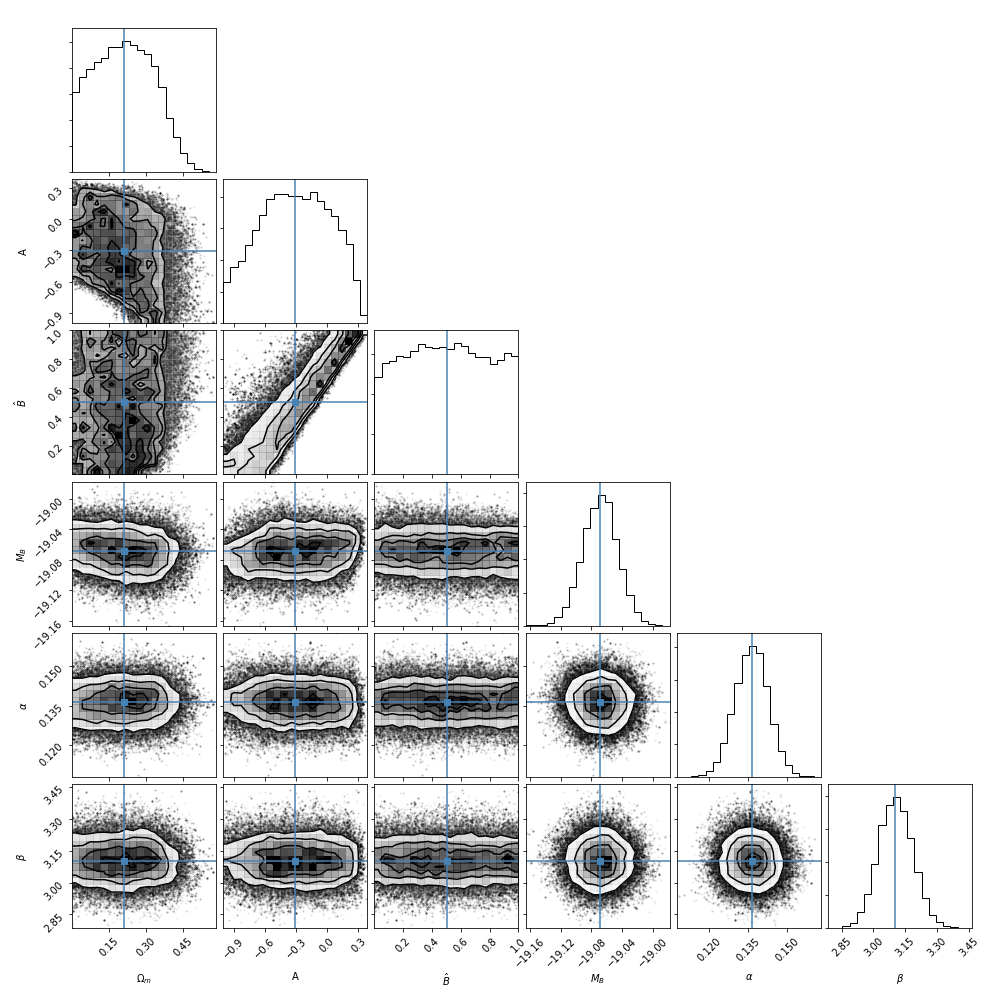} 
\caption{\label{triangle2}Joint and marginalized constraint of $\Omega_{m}$, $A$ and $\hat{B}$, for the model presented in the second solution, and marginalized constraint of the light-curve parameters $M_{B}$, $\alpha$ and $\beta$ of the JLA sample. The admissible regions correspond to $1\sigma(68.3\%)$, $2\sigma(95.5\%)$ and $3\sigma(99.7\%)$ confidence level (CL), respectively. The best fit values of each parameter are shown in Table \ref{Table I}.}
\end{figure*}

\begin{figure*}
\centering
\includegraphics[scale=0.50]{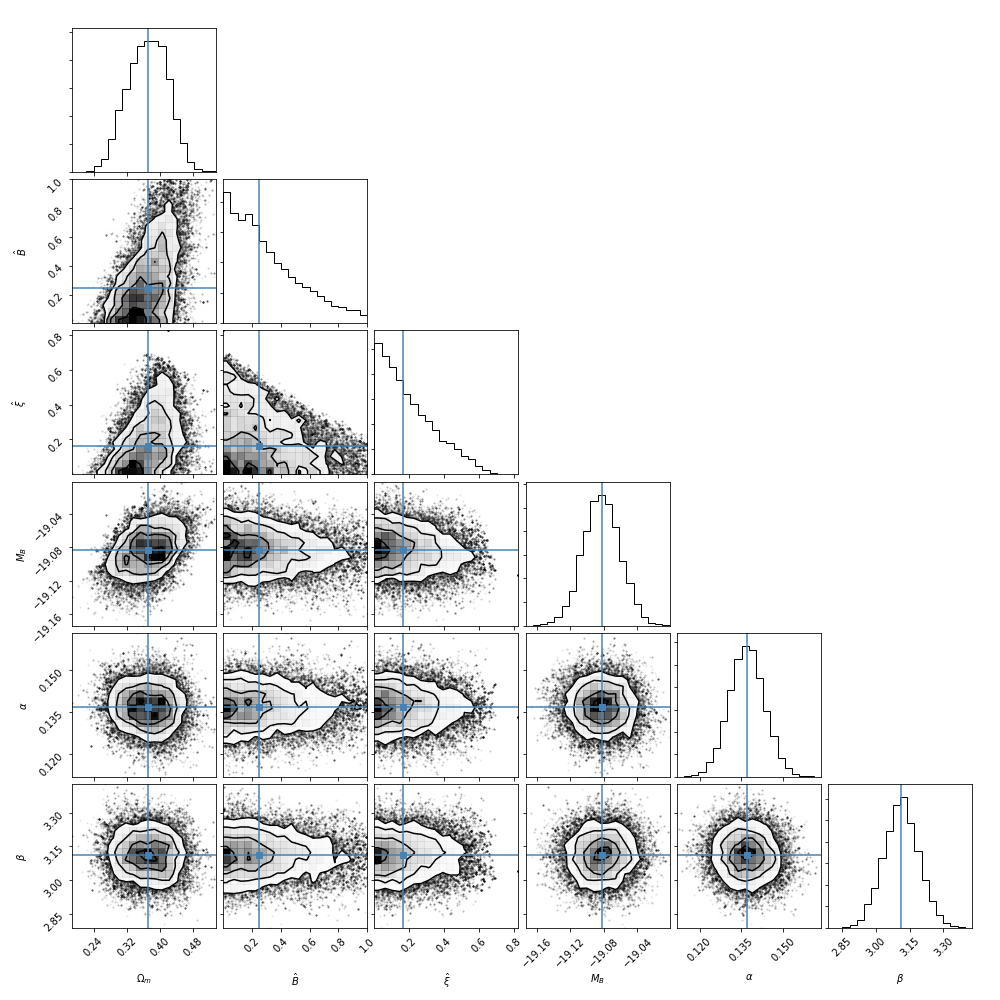} 
\caption{\label{triangle3}Joint and marginalized constraint of $\Omega_{m}$, $\hat{B}$ and $\hat{\xi}$, for the model presented in the third solution, and marginalized constraint of the light-curve parameters $M_{B}$, $\alpha$ and $\beta$ of the JLA sample. The admissible regions correspond to $1\sigma(68.3\%)$, $2\sigma(95.5\%)$ and $3\sigma(99.7\%)$ confidence level (CL), respectively. The best fit values of each parameter are shown in Table \ref{Table I}.}
\end{figure*}

The theoretical distance modulus of SNe is defined as
\begin{equation}
\mu_{th}\left(z,\vec{p}\right)=5\log_{10}{\left[\frac{d_{L}\left(z,\vec{p}\right)}{Mpc}\right]}+25, \label{mutheo}
\end{equation}
where the vector $\vec{p}$ collects the parameters, i.e., it is built with the free parameters of each theoretical model, and $d_{L}$ is the luminosity distance, given by
\begin{equation}
d_{L}\left(z,\vec{p}\right)=\frac{c\left(1+z\right)}{H_{0}}\int_{0}^{z}{\dfrac{dz'}{E\left(z',\vec{p}\right)}}, \label{lumdistance}
\end{equation}
where $c$ is the speed of light given in units of $km/s$, $H_{0}$ is the current Hubble parameter for which we consider the fixed fiducial value of $70[km\: s^{-1}/Mpc]$ and $E\left(z,\vec{p}\right)$ is defined by
\begin{equation}
H\left(z,\vec{p}\right)=H_{0}E\left(z,\vec{p}\right). \label{defofE}
\end{equation}

On the other hand, in the JLA sample the distance estimator used assumes that supernovae with identical color, shape and galactic environment have on average the same intrinsic luminosity for all redshift. This hypothesis is quantified by a linear model as
\begin{equation}
\mu=m_{b}^{*}-\left(M_{B}-\alpha\times X_{1}+\beta\times C\right), \label{muobserv}
\end{equation}
where $m_{b}^{*}$ correspond to the observed peak magnitude in rest frame $B$ band, $X_{1}$ is the stretch parameter, $C$ is the color parameter and $M_{B}$, $\alpha$ and $\beta$ are nuisance parameters in the distance estimate. So, these last three parameters have to be computed and marginalized simultaneously with the free parameters present in the vector $\vec{p}$.

To compute the best-fit parameters we use the Affine Invariant Markov Chain Monte Carlo (MCMC) method~\cite{Goodman}, implemented  in the pure-Phyton code \textit{emcee}~\cite{Foreman} with a likelihood given by the following Gaussian distribution
\begin{equation}
\mathcal{L}=\mathcal{N}e^{-\chi^{2}/2}, \label{likelihood}
\end{equation}
where $\mathcal{N}$ is a normalization constant. Following~\cite{Betoule}, the distance estimate of Eq.(\ref{muobserv}) can be written in matrix notation, by forming a matrix $\mathbf{A}$ such that
\begin{equation}
\bm\mu=\mathbf{A}\bm\eta-\mathbf{M_{B}}, \label{mumatrix}
\end{equation}
where
\begin{equation}
\bm\eta=\left(\left(m_{b,1}^{*}, X_{1,1}, C_{1}\right),\ldots ,\left(m_{b,n}^{*}, X_{1,n}, C_{n}\right) \right), \label{etavector}
\end{equation}
\begin{equation}
\mathbf{A}=\mathbf{A_{0}}+\alpha\mathbf{A_{1}}-\beta\mathbf{A_{2}}, \; \textup{with} \; \left(\mathbf{A_{k}}\right)_{i,j}=\delta_{3,j+k}, \label{Amatrix}
\end{equation}
are the $n$-dimensional vector and the $n\times n$ matrix respectively, with $n=740$ the number of SNe samples. Also, the JLA sample provides a covariance matrix $\mathbf{C}$, which encodes the statistical and systematic uncertainties. Hence, the $\chi^{2}$ function of Eq.(\ref{likelihood}) has the form
\begin{equation}
\chi^{2}=\left(\bm\mu\left(\vec{p}_{J}\right)-\bm\mu_{th}\left(z,\vec{p}\right)\right)^{\dagger}\mathbf{C}^{-1}\left(\bm\mu\left(\vec{p}_{J}\right)-\bm\mu_{th}\left(z,\vec{p}\right)\right), \label{defofchi}
\end{equation}
where $\vec{p}_{J}=\left(M_{B},\alpha,\beta\right)$. This is the expression for $\chi^{2}$ that we will use in our MCMC analyses, a function that will be minimized in order to compute the best-fit values and confidence intervals. In this procedure, we use for the vector parameters $\vec{p}_{J}$ the following priors: $-20<M_{b}<-18$, $0<\alpha<1$ and $0<\beta<5$.

It is necessary to emphasize that the corresponding scale factor of each solution is only valid when the matter density can be neglect in comparison to the dark energy one. So, for the fit we use Eqs.(\ref{Fr1}) and (\ref{defofE}), adding the usual matter component, thus
\begin{equation}
E(z,\vec{p})=\sqrt{\frac{\rho_{m}}{3H_{0}^{2}}+\frac{\rho_{DE}}{3H_{0}^{2}}}, \label{Eforfit}
\end{equation}
where $\rho_{m}/3H_{0}^{2}$ is the matter component, given by the expression
\begin{equation}
\frac{\rho_{m}}{3H_{0}^{2}}=\Omega_{m}\left(1+z\right)^{3}, \label{rhomatter}
\end{equation}
and $\rho_{DE}/3H_{0}^{2}$ is the dark energy component, which depends of each model, as follows:

\textbf{i)} Considering the relation between the scale factor $a$ and the redshift $z$, given by the expression $a=\left(1+z\right)^{-1}$, then the conservation equation (\ref{ct}) can be written as
\begin{equation}
\dfrac{d\rho}{dz}-\frac{3}{\left(1+z\right)}\left(\rho+P\right)=0. \label{ctredshift}
\end{equation}

\textbf{ii)} For the first solution, we consider the equation of state (\ref{7}) in order to solve the first-order differential equation (\ref{ctredshift}) with the initial condition $\rho(z=0)=\rho_{0}$, obtaining the following expression
\begin{equation}
\rho(z)=\left[\rho_{0}^{1/2}-\frac{3B}{2}\ln{\left(1+z\right)} \right]^{2}. \label{rhofirst}
\end{equation}
Thus, if we consider that $\rho(z)=\rho_{DE}(z)$, then the dark energy component of Eq.(\ref{Eforfit}) is given by
\begin{equation}
\frac{\rho_{DE}(z)}{3H_{0}^{2}}=\left[\left(1-\Omega_{m}\right)^{1/2}-\hat{B}\ln{\left(1+z\right)} \right]^{2}, \label{rhoDEfirst}
\end{equation}
where we have introduced  the dimensionless constants
\begin{equation}
\hat{B}=\frac{\sqrt{3}B}{2H_{0}} \;\; \textup{and} \;\; \Omega_{DE}=\frac{\rho_{0}}{3H_{0}^{2}}. \label{constantfirst}
\end{equation}
In Eq(\ref{rhoDEfirst}) the constraint $\Omega_{m}+\Omega_{DE}=1$ was used, therefore the vector parameters of this model is $\vec{p}=\left(\Omega_{m},\hat{B}\right)$ for which we use the priors: $0<\Omega_{m}<1$ and $0<\hat{B}<1$.

\textbf{iii)} For the second solution, the expression for the energy density, using the equation of state (\ref{2}) with $A>-1$ and $B>0$, is given by
\begin{equation}
\begin{split}
& \rho(z)=\frac{1}{\left(1+A\right)^{2}}\times \\
& \left\{\left[\left(1+A\right)\rho_{0}^{1/2}-B\right]\left(1+z\right)^{3(1+A)/2}+B\right\}^{2}. \label{rhosecond}
\end{split}
\end{equation}
Thus, the dark energy component of Eq.(\ref{Eforfit}) reads
\begin{equation}
\begin{split}
& \frac{\rho_{DE}(z)}{3H_{0}^{2}}=\frac{1}{\left(1+A\right)^{2}}\times \\
& \left\{\left[\left(1+A\right)\left(1-\Omega_{m}\right)^{1/2}-\hat{B}\right]\left(1+z\right)^{3(1+A)/2}+\hat{B}\right\}^{2}, \label{rhoDEsecond}
\end{split}
\end{equation}
where we define the dimensionless constants (observe that $A$ is already dimensionless)
\begin{equation}
\hat{B}=\frac{B}{\sqrt{3}H_{0}} \;\; \textup{and} \;\; \Omega_{DE}=\frac{\rho_{0}}{3H_{0}^{2}}. \label{constantsecond}
\end{equation}
Again, the constraint $\Omega_{m}+\Omega_{DE}=1$ was used, therefore the vector parameters of this model are $\vec{p}=\left(\Omega_{m},A,\hat{B}\right)$ for which we used the priors: $0<\Omega_{m}<1$, $-1<A<1$ and $0<\hat{B}<1$.

\textbf{iv)} For the third solution, it is not possible to obtain an analytic expression for the energy density as a function of the redshift, because Eq.(\ref{33}) is not an injective function. Therefore, in this case, and for simplicity, we have used Eqs.~(\ref{34}) and (\ref{35}) in order to obtain numerically $\rho$ as a function of $a$ and, consequently, as a function of $z$. As initial condition, we used $t_{0}=0$, $a(t=0)=1$, and we have defined the following dimensionless constants
\begin{equation}
\hat{B}=\frac{\sqrt{3}B}{H_{0}}, \;\; \hat{\xi}=\frac{\xi}{H_{0}^{2}} \;\; \textup{and} \;\; \Omega_{DE}=\frac{\rho_{0}}{3H_{0}^{2}}. \label{constantthird}
\end{equation}
Again, the constraint $\Omega_{m}+\Omega_{DE}=1$ was employed, so that the vector with the parameters of this model reads $\vec{p}=\left(\Omega_{m},\hat{B},\hat{\xi}\right)$, for which we used the priors: $0<\Omega_{m}<1$, $0<\hat{B}<1$ and $0<\hat{\xi}<1$. It is important to mention that, for some values of $\Omega_{m}$, $\hat{B}$ and $\hat{\xi}$, the bouncing could occur near $z=0$, i. e. at present time. Therefore, we need to impose the physical requirement that the bounce occurs for, at least,  $z>1.4$. This last condition actually eliminates the double value behavior of the solution at the redshift coming from the JLA sample.

\section{Results and discussion}
\begin{table*}
\centering
\begin{tabular}{|l|lllllll|lll|}
\hline
\multicolumn{1}{|c|}{Model} & \multicolumn{7}{c|}{Best fit values} & \multicolumn{3}{c|}{Goodness of fit} \\
\hline\hline
 & \multicolumn{1}{c}{$\Omega_{m}$} & \multicolumn{1}{c}{$A$} & \multicolumn{1}{c}{$\hat{B}$} & \multicolumn{1}{c}{$\hat{\xi}$} & \multicolumn{1}{c}{$M_{B}$} & \multicolumn{1}{c}{$\alpha$} & \multicolumn{1}{c|}{$\beta$} & $\chi^{2}_{min}$ & AIC & BIC \\
\hline
$\Lambda$CDM & $0.292_{-0.032}^{+0.035}$ & \multicolumn{1}{c}{-} & \multicolumn{1}{c}{-} & \multicolumn{1}{c}{-} & $-19.076_{-0.021}^{+0.022}$ & $0.137_{-0.006}^{+0.006}$ & $3.108_{-0.081}^{+0.083}$ & $692.1$ & $700.1$ & $718.5$ \\
\hline
First solution & $0.347_{-0.049}^{+0.062}$ & \multicolumn{1}{c}{-} & $0.18_{-0.13}^{+0.23}$ & \multicolumn{1}{c}{-} & $-19.082_{-0.022}^{+0.021}$ & $0.137_{-0.007}^{+0.007}$ & $3.11_{-0.08}^{+0.08}$ & $693.1$ & $703.1$ & $726.1$ \\
\hline
Second solution & $0.21_{-0.13}^{+0.13}$ & $-0.307_{-0.365}^{+0.362}$ & $0.503_{-0.326}^{+0.334}$ & \multicolumn{1}{c}{-} & $-19.068_{-0.023}^{+0.022}$ & $0.136_{-0.006}^{+0.006}$ & $3.104_{-0.078}^{+0.081}$ & $692.1$ & $704.1$ & $731.7$  \\
\hline
Third solution & $0.372_{-0.052}^{+0.047}$ & \multicolumn{1}{c}{-} & $0.249_{-0.178}^{+0.317}$ & $0.165_{-0.119}^{+0.203}$ & $-19.083_{-0.022}^{+0.022}$ & $0.137_{-0.006}^{+0.006}$ & $3.112_{-0.077}^{+0.077}$ & $693.8$ & $705.8$ & $733.4$ \\
\hline
\end{tabular}
\caption{Best fit values for each model parameters, $\vec{p}$, as well as the respective goodness of fit criteria and light-curve parameters, $\vec{p}_{J}$, of the JLA sample. The first row shows the best fit values for the standard cosmological model, $\Lambda$CDM; the second, third and fourth rows correspond to the best fit parameters for the first, second and third solutions, respectively, which were analyzed in Section III as a regular and Little Rip solutions. We have focused on the Bayesian criterion information in order to determine the best model to fit the data, and to compare the solutions with the $\Lambda$CDM model.}
\label{Table I}
\end{table*}

All our solutions will be compared with the $\Lambda$CDM model, whose respective $E(z,\vec{p})$ is given by
\begin{equation}
E_{\Lambda CDM}(z,\vec{p})=\sqrt{\Omega_{m}\left(1+z\right)^{3}+\Omega_{\Lambda}}, \label{ELambdaCDM}
\end{equation}
where $\Omega_{\Lambda}=1-\Omega_{m}$, i.e. the vector of parameters of this model is again given by $\vec{p}=(\Omega_{m})$. In order to compare the goodness of the fits, we will use the Akaike Information Criterion (AIC), which is defined as
\begin{equation}
AIC=2k-2\ln{\left(\mathcal{L}_{max}\right)}, \label{AIC}
\end{equation}
where $\mathcal{L}_{max}$ is the maximum value of the likelihood function, calculated for the best fit parameters, $n$ is again  the number of SNe sample and $k$ the number of free parameters of the model. In addition, we also calculate the Bayesian Criterion Information, defined as
\begin{equation}
BIC= k\ln{\left(n\right)}-2\ln{\left(\mathcal{L}_{max}\right)}. \label{BIC}
\end{equation}
Both the AIC and BIC criteria try to solve the problem of maximizing the likelihood function by adding free parameters, resulting in overfitting. To resolve this problem both criteria introduce a penalization that depends on the total number of free parameters of our model, which is higher in the BIC case that in the AIC case, because the penalization in the first one depends on the natural logarithm of the total observational data. The model favored by  observations, as compared to the other, corresponds to the one with the smallest value of AIC/BIC. Hence, we focused our analysis on the BIC criterion, where in general a difference of $2-6$ in BIC between the two models is considered as an evidence against the model with the higher BIC, a difference of $6-10$ in BIC  is already a strong evidence, and a difference $>10$ in BIC is definitely a very strong evidence.

The best fit values for each model as well as the goodness of fit criterion are show in Table \ref{Table I}. In Figs.~\ref{trianglelcdm}-\ref{triangle3} we depict the joint credible regions of the $\Lambda$CDM model and the first three Little Rip solutions studied here, for combinations of their respective vectors of parameters $\vec{p}$ and $\vec{p}_{J}$.

As we can see in Table \ref{Table I}, the $\Lambda$CDM model and the three solutions that we have tested exhibit a very similar goodness of fit, as given by the value of $\chi_{min}^{2}$. Even more, the second solution has exactly the same value of $\chi_{min}^{2}$ than the $\Lambda$CDM model, and the first and second solution  differ only in $1$ and $1.7$, respectively, in the value of $\chi_{min}^{2}$, in relation with the $\Lambda$CDM model. But, the AIC criterion does show us that, statistically, the $\Lambda$CDM model is the best one, because it has the minimum value for $AIC$. Even more, the difference in the AIC between the $\Lambda$CDM and the second solution is $4$, i. e., these two models fit well the supernovae data but the first one is in fact better than the second. The BIC criterion leads to this conclusion more clearly. Again, the lowest value of BIC correspond to the $\Lambda$CDM model, followed by the first solution, whose value of BIC differs from the one for $\Lambda$CDM in $7.6$. Thus, in this case we have strong evidence against the first solution. For the second and third solutions, we have a very strong evidence against them.

\begin{figure}[ht]
\includegraphics[width=20pc]{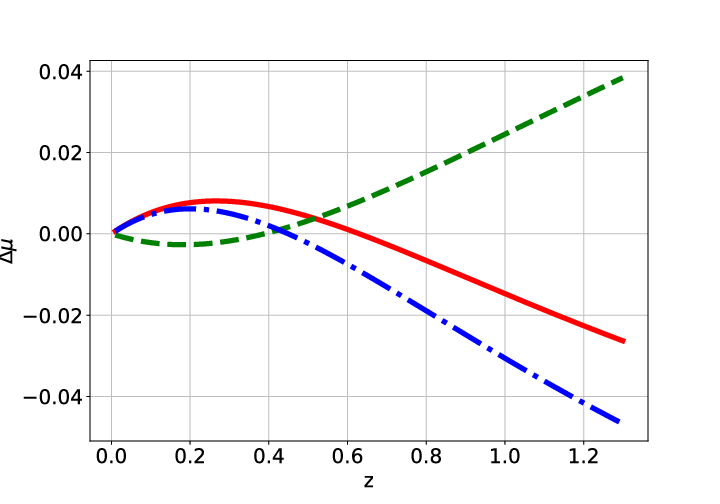}
\caption{\label{Dmu}Plot of the distance modulus for $\Lambda$CDM model subtracted from the distance modulus for the first solution (line), second solution (dashed) and the third solution (dashed-dotted). By definition the $\Lambda$CDM model is represented by $\Delta\mu =0$.}
\end{figure}

The important point here is that the three solutions tested above have higher values in the AIC/BIC tests than the $\Lambda$CDM one, because they have more free parameters. But, these extra free parameters are added only in order to obtain a phantom dark energy and not for improving the fit with the supernova data. Thus, we conclude that these three models, which represent a phantom dark energy with a Little Rip behavior at late time and regular behavior at early time, do fit well the supernova data. Even more, in Fig.(\ref{Dmu}) we see that these three models just differ very slightly from the $\Lambda$CDM model, albeit in essence they are actually very different.

Finally, if we focuss on the values of $\Omega_{m}$ for the first and third solutions, we observe that these values remain in the acceptance region for the value $\Omega_{m}=0.315\pm 0.007$ given by the latest Planck~\cite{Planck} survey data; but they can reach values of $0.4$ and more. A possible explanation of this fact is that the free parameters added to the cosmological constant (in these two cases the EoS is a deviation to the phantom region of the cosmological constant) allow to obtain an accelerated universe at present time with a lower value of the dark energy component (see, for example,~\cite{Gonzalez}).

\section{Conclusions and final remarks}
We have shown in this paper that some previously considered cosmological solutions for a flat universe filled with a GEoS with a phantom behavior can actually yield regular solutions at late times, avoiding Big Rip singularities and fitting well the supernova data, what makes of them viable models at late times. Additionally, they can also give rise to early time regular solutions, like emergent or bouncing universes, without an initial singularity. In other words, these solutions are regular ones for all time, except at the (unreachable) asymptotic limit $t\rightarrow\infty$, provided some very reasonable conditions are fulfilled.

We  have also proved that, for the bouncing models of flat FLRW metric in a phantom regime, the EoS parameter $\omega$ associated to the EoS converges to minus infinity at the time of the bounce, namely that $\omega\left(t_{b}\right)=-\infty$. This means that, although the bouncing universes are regular solutions for the flat FLRW universe, their EoS parameter cannot be defined at the time of bounce as a function of the cosmological time.

The condition to avoid Big Rip solutions was found in~\cite{Frampton}. We have here extended this criteria in order to find the conditions that allow to avoid the initial singularity and the late time singularity of Big Rip type simultaneously. This is a most remarkable result. We have shown that in a flat space dominated by a fluid given by a GEoS with an EoS parameter  $\omega<-1$ in the FLRW metric, only the Bouncing and Emergent universes are free from singularities, and that all solutions with a scale factor that can be written in the form $e^{g(t)}+s$ will represent  Regular and  Little Rip Universes if and only if $g(t)$ satisfies the conditions (\ref{crit1}) for $s=0$ and (\ref{crit2}) for $s>0$.

Using those conditions, we have investigated five different solutions, which had been previously discussed in the literature, but always in the context of either their late-time or early-time behavior only,  never  in both domains consistently. The link that can be established, by means of the above conditions, between the regular solutions occurring at early and late times, respectively, has proven to be very powerful in extending the procedure to get regular solutions valid in both regions, simultaneously. The final result has been, in each case, to produce new cosmological solutions that are non-singular for all finite time, a considerable extension of the family of regular solutions that had been found previously in the literature.

It is worth mentioning that the phantom behavior of all the solutions considered is a key feature of the method; indeed, this allows to establish the link found here.  The result, from the physical point of view, is that we now have a well grounded theoretical model, which explains the phantom behavior consistently, and which opens the possibility for a solid description of the early and late time stages of the universe in a consistent way, without singularities. If the results of the latest astronomical surveys, which point towards a phantom cosmology, are confirmed by more precise observations, the importance of the theoretical models here obtained might be paramount.

\begin{acknowledgments}
This work was supported by CONICYT through Grant FONDECYT N$^{\circ}$Y 1110840 (N.C.) and CONICYT-PCHA/Doctorado Nacional/2016-21160331 (E.G.). EE and SDO are partially supported by MINECO (Spain), Project FIS2016-76363-P, by AGAUR (Generalitat de Catalunya), Project 2017SGR247, and by the CPAN Consolider Ingenio Project. EE was partially supported by the Yukawa Institute for Theoretical Physics at Kyoto University, where part of this work was done during the workshop YITP-T-17-02 ``Gravity and Cosmology 2018''. E. G. and N. C. acknowledge Dr. Arturo Avelino from Harvard-Smithsonian Center for Astrophysics, for his help in understanding the supernova data fit.
\end{acknowledgments}

\appendix
\section{Proof of non-singularity for $\omega <-1/3$}
In this Appendix we  prove that in a flat space dominated by a fluid, with  EoS parameter  $\omega <-1/3$, in the FLRW metric,  the only models without singularities are the Bouncing and the Emergent ones.

\begin{figure}[ht]
\centering
\subfigure[Plot of the scale factor as a function of time.]{\includegraphics[width=18pc]{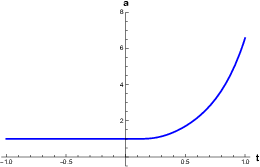}}
\subfigure[Plot of the parameter of state as a function of time.]{\includegraphics[width=18pc]{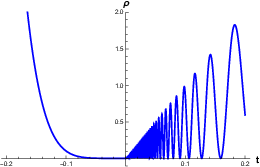}}
\caption{Examples}
\label{Ej}
\end{figure}

The definition of the EoS parameter  $\omega $ in Eq.~(\ref{parestado}) is used  not to discard the case $P(t_{b})=\rho(t_{b})=0$. Furthermore,  $\omega$ is required to be defined for all time, except perhaps at a point where it tends to  $-\infty $. In this way, it is not allowed that $\rho$ can be zero in a whole interval, even if $P$ is also zero. The next example illustrates a function $\omega(t)$ that is not well defined for all $t$. Consider an emergent solution with smooth functions $\rho$, $P$ and $a$, obtained from a scale factor as an Ansatz. The behavior of the scale factor is constant for $t\leq 0$ and exponential for $t>0$, with $\omega<-1$ for $t>0$. This behavior is represented in Fig.~\ref{Ej}.a.

As we can see, this model is an emergent universe of the Little Rip type, where $a$, $\rho$ and $P$ are of class $C^{\infty}$ and the scale factor is a convex function, thus $\ddot{a}\geq 0$. In this case,  $\omega<-1$ for all $t>0$, but it is not defined for values $t\leq 0$, since $P(t)=\rho(t)=0$ for $t\leq 0$. Thus, $\rho $ cannot be $0$ in a whole interval (and therefore neither $\dot{a}$). Furthermore, if we demand that $\rho\in C^{\infty }$, then the set of points $T$ such that $\rho(t)=0$  must be numerable and the adherence of any subset $T_{i}$ of this cannot be equal to the interval $[x,y]$, for any $x,y\in \mathbb{R}$. Otherwise, there would be intervals in which the function $\rho$ would jump from the points where it is nonzero to zero discontinuously. Therefore, at most a numerable amount of points $t_{n}$ may be considered such that $\rho (t_{n})=0$.

The next is an example that yields the existence of functions $\rho(t)$ with a numerable  infinite quantity of points where $\rho(t)=0$, with $\omega$ well defined at all points, and with the function $\rho \in C^{\infty}$. This is a valid solution of the Friedmann equations and one that avoids any kind of singularity, yielding an $\omega$ well defined for all time, but one that does not satisfy the condition  $\omega<-1/3$. Consider the energy density as an increasing oscillating function for $t>0$ and with exponential behavior for $t<0$, with $\rho (0)=0$. This behavior is represented in Fig.~\ref{Ej}.b. Without further difficulties, it is possible to see that, then, $\omega $ is well defined at all points, and that $\omega<-1/3$ in some piece of each cycle.

After  the above preamble, let us continue with the proof. Suppose $\omega<-\frac{1}{3}$, and that the set of points $T$ is such that $\rho(t)=0$ is numerable, and the adherence of any subset $T_{i}$ of $T$ is not equal to the interval $[x,y]$ for any $x,y\in \mathbb{R}$. Then, there exits at most one point $t$ such that $\rho(t)=0$. Indeed, suppose there would be a pair of points $t_{a}<t_{b}$ with $\rho(t_{a})=\rho(t_{b})=0$. By hypothesis, it is possible to consider that in the interval $I=(t_{a},t_{b})$ there is no other point fulfilling the condition $\rho(t)=0$. As $\rho$ is defined to be non-negative ($\rho=3H^{2}$) and $\omega <-\frac{1}{3}$, then $\ddot{a}(t)>0$, for any point in $I$, since $\rho+3P=\rho(1+3\omega )<0$. Then, the function $\dot{a}$ is strictly increasing in $I$. On the other hand, $\rho(t_{a})=0$ so that $\dot{a}(t_{a})=0$. Thus, since $\dot{a}$ is strictly increasing, it turns out that $\dot{a}(t_{a})<\dot{a}(t_{b})$, therefore $\dot{a}(t_{b})>0$, and then $\rho (t_{b})>0$. But this is a contradiction, because it is supposed that $\rho (t_{b})=0$. As a consequence, not more than one point $t$ can exist such that $\rho (t)=0$.

Finally, if $\rho$ is equal to $0$ at most at one point, then $H$ is also  equal to $0$ at most in that point, and therefore, $\dot{a}$ too. This last fact and the convexity of $a$ imply that the only valid models of the Friedmann equation that avoid the initial singularity are the  Bouncing and Emergent models.

\section{Proof of the criterion for Bouncing universes}
In this Appendix we will prove a property that is used in Sect. II.C.

If $f$ is an analytic function at $t_{0}$ such that $f(t_{0})=\dot{f}(t_{0})=0$, then $\lim _{t\rightarrow t_{0}} \left|\dfrac{\dot{f}(t)}{f(t)}\right|= \infty $. Indeed, without loss of generality, it is possible to consider $t_{0}=0$, since it is enough to define $g(t)=f(t+t_{0})$. Now, as $f$ and $\dot{f}$ are analytic functions, it is possible to represent them in terms of their  Maclaurin series, namely
\begin{equation}
\begin{array}{l}
f(t)=\sum_{n=0}^{\infty }\alpha _{n}t^{n},\\ \\
\dot{f}(t)=\sum _{n=0}^{\infty}\alpha _{n+1}(n+1)t^{n}.
\end{array}
\end{equation}
Let $m\in\mathbb{N}$ be the first integer such that $\alpha_{m}\neq 0$. By assumption, $f(0)=\dot{f}(0)=0$, therefore $m\geq 2$. Then the Maclaurin series of $f$ and $\dot{f}$  are represented by
\begin{equation}
\begin{array}{l}
f(t)=t^{m}\sum_{n=0}^{\infty }\alpha _{n+m}t^{n},\\ \\
\dot{f}(t)=t^{m-1}\sum _{n=0}^{\infty}\alpha _{m+n}(m+n)t^{n},
\end{array}
\end{equation}
respectively.
Finally, applying the limit to $\left|\frac{\dot{f}}{f}\right|$, we reach the announced conclusion, namely that
\begin{widetext}
\begin{equation}
\lim _{t\rightarrow 0}\left|\dfrac{\dot{f}(t)}{f(t)}\right|=\lim _{t\rightarrow 0} \left|\dfrac{t^{m-1}\sum _{n=0}^{\infty}\alpha _{m+n}(m+n)t^{n}}{t^{m}\sum_{n=0}^{\infty }\alpha _{n+m}t^{n}}\right|= \lim _{t\rightarrow 0} \dfrac{1}{|t|}\dfrac{m|\alpha _{m}|}{|\alpha_{m}|}=\lim _{t\rightarrow 0} \dfrac{m}{|t|}=\infty.
\end{equation}
\end{widetext}

\end{document}